\documentclass[prl,twocolumn,letterpaper,showpacs]{revtex4}

\usepackage[dvips]{graphicx}
\usepackage{dcolumn}
\usepackage{bm}
\usepackage{xspace}
\usepackage{amssymb}
\countdef\preprint=0 \countdef\tube=12

\begin{document}

\bibliographystyle{aip}

\title{Optical Measurement of the Phase-Breaking Length  in Graphene}

\author{Luiz Gustavo Can\c{c}ado, Ryan Beams, and Lukas Novotny.}
\homepage{http://www.nano-optics.org}

\affiliation{Institute of Optics, University of Rochester, Rochester NY (USA)}

\date{\today}

\begin{abstract}
This paper reports the experimental determination of the {\em phase-breaking length} $L_{\phi}$ of conduction electrons in graphene using Raman spectroscopy. Based on the double-resonance model, we extract $L_{\phi}$ from the spatial dependence of the $D$ band susceptibility $\chi_D$ near the graphene edge. By using prior knowledge of sample properties and the excitation point-spread function we are able to determine the spatial variation of the Raman susceptibilities with high accuracy, and the results reveal a {\em phase-breaking length} $L_{\phi}$\,=\,40\,nm near the graphene edge.
\end{abstract}

\pacs{78.67.-n, 73.63.-b, 63.20.kr, 78.30.-j}

\maketitle

The unique electronic properties of graphene are attracting considerable scientific and technological interest~\cite{geim}. Conduction electrons in graphene behave as massless and relativistic two-dimensional Dirac fermions that give rise to an unusual quantum hall effect~\cite{zhang,novo01,novo02}. The very high electron mobility at room temperature detected in micron-sized graphene samples (5000-10000\,cm$^{2}$V$^{-1}$s$^{-1}$)\cite{novo04,mean01,berger02} is already being explored for novel ballistic-transport devices, including graphene-based field-effect transistors (FETs) and single-electron transistors (SETs)~\cite{berger,geim}. However,  the physical properties of nanoscale graphene devices are predicted to be strongly affected by the edges, which act as defects in graphene's crystal structure~\cite{louis,denis}. Recent studies of the transport properties of graphene nanoribbon based FET architectures revealed the strong impact of edge states for nanoribbons having widths below 50nm~\cite{Avouris}. Therefore, it is important to understand the symmetry breaking nature of the edges and associated local electronic effects.

The Raman spectrum of graphite is known to exhibit several peaks that are only observed in the presence of structural defects in the hexagonal lattice~\cite{tan,thomrev,pimenta}. Of particular interest is the $D$ band ($\sim$\,1350cm\,$^{-1}$) that appears close to the first-order allowed peak denoted as $G$ band ($\sim$\,1580\,cm$^{-1}$)~\cite{tuinstra01}. Electrons involved in the double-resonance process giving rise to the $D$ band are inelastically scattered by phonons in the interior of the first Brillouin zone. However, because of momentum conservation these phonons can only become Raman active if the electrons involved in the process are elastically scattered by a defect~\cite{thomsen,saito}. The intensity ratio between the disorder-induced $D$ band and the $G$ band is widely used as a measure for the average crystallite size $L_{a}$ in graphitic systems~\cite{tuinstra01}. Single layers, bilayers and few layers of graphene can be identified by the lineshape of the $G^{\prime}$ band (the second harmonic of the $D$ band occurring at $\sim$\,2700\,cm$^{-1}$)~\cite{ferrari02,eklund,graf}. Recent Raman scattering studies have also revealed that the electron-phonon interaction in graphene can be affected by electric fields~\cite{castro,ferrari01,yan} and that the adiabatic Born-Oppenheimer approximation can be violated~\cite{ferrari01}.

In this paper, we use confocal Raman spectroscopy to experimentally determine the {\em phase-breaking length} $L_{\phi}$ of conduction electrons in graphene. $L_{\phi}$ is the average distance traveled by an electron before undergoing inelastic scattering with a lattice phonon~\cite{solid01}. We extract $L_{\phi}$ from the spatial dependence of the $D$ band susceptibility $\chi_D$ near the graphene edge. Based on the double-resonance model, $\chi_D$ is directly associated with inelastic scattering of transverse optical phonons near the vertices of the $1^{st}$ Brillouin zone of graphene.
Magnetotransport measurements predict a {\em phase-breaking length} $L_{\phi}\,\approx$\,100\,nm for micron-sized graphene samples at room temperature~\cite{mean01,berger02}. To the best of our knowledge, a direct optical measurement has not been performed so far.

At first sight, length scales on the order of $L_{\phi}$ cannot be resolved with a diffraction-limited confocal imaging system. However, using prior knowledge of sample properties (edge position) and the excitation (point-spread function) we are able to determine the spatial variation of the Raman susceptibilities with an accuracy of $\approx 10\,$nm. The confocal measurements are further guided by simultaneous topographic imaging with a shear-force atomic-force microscope (AFM)~\cite{shear}, and the results reveal a {\em phase-breaking length} of 40\,nm near the graphene edge. This value is shorter than that predicted for micron-sized graphene samples at room temperature~\cite{mean01,berger02}, and shows the influence of the edges on the transport properties of charge carriers in nanoscale graphene systems (such as nanoribbons), where edge defects play an important role~\cite{louis,denis}.

The graphene samples used in this study were prepared by micromechanical cleavage of a piece of highly oriented pyrolytic graphite at the surface of a clean microscope cover glass, following the method reported in reference \cite{novo03}. Single layer graphene flakes were identified by optical microscopy and subsequent topographic characterization with shear-force microscopy using a chemically etched glass tip attached to a quartz tuning fork. The confocal Raman instrument used in this experiment is based on an inverted optical microscope equipped with an x,y-scan stage. A high numerical aperture objective (1.4\,NA) is used to focus a linearly polarized laser beam with wavelength 632.8\,nm on the sample surface. The scattered light is collected using the same objective and then detected with a single-photon counting avalanche photodiode (APD) or a combination of a spectrograph and a cooled charge-coupled device (CCD).

\begin{figure}
\begin{center}
\includegraphics[width=0.38\textwidth]{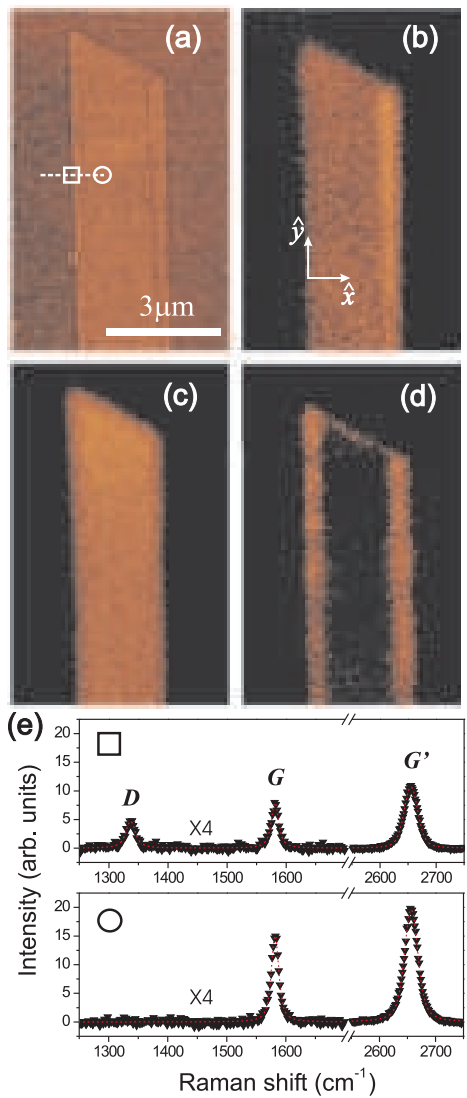}
\caption{(a): Topographic image of a graphene layer on a glass substrate. (b)-(d): Confocal Raman images showing the $G$, $G^{\prime}$ and $D$ band intensities over the same region. (e): Raman spectra of a single graphene layer. Top: spectrum recorded near the edge [white square in (a)]; bottom: spectrum recorded in the interior [white circle in (a)]. The spectra are scaled by a factor of 4 in the range $1250-1700\,cm^{-1}$ .\label{fig1}}
\end{center}
\end{figure}

Figure \ref{fig1}(a) shows the topographic image of a single graphene layer on a glass substrate. Figures \ref{fig1}(b)-(d) are corresponding confocal Raman images showing the $G$, $G^{\prime}$ and $D$ band intensities, respectively. The $G$ band is a first-order scattering process that originates from the double degenerate vibrational mode $\Gamma_{6}^{+}$ (E$_{2g}$) that occurs at the crossing of the longitudinal optical (LO) and transverse optical (TO) phonon branches at the $\Gamma$ point in the 1$^{st}$ Brillouin zone of graphene \cite{tuinstra01}. Notice that the $G$ band intensity is roughly uniform along the graphene surface, as shown in Figure \ref{fig1}(b). A similar situation occurs for the $G^{\,\prime}$ band, which is the overtone of the $D$ band but does not require a disorder-induced process to become Raman active, since momentum conservation is guaranteed in two-phonon Raman processes \cite{stokes}. On the other hand, the $D$ band can be detected only near the graphene edges, which act as defects necessary for momentum conservation in the one-phonon double-resonance process involving phonons in the interior of the $1^{st}$ Brillouin zone \cite{edge}.

All confocal Raman images shown in Figures \ref{fig1} were recorded with the polarization vector ($\vec{P}_{0}$) of the excitation laser beam oriented parallel to the graphene edge [$y$-direction in Fig.~\ref{fig1}(b)]. Polarizations perpendicular to the edge do not generate any $D$ band Raman scattering~\cite{edge}. This is the reason why the $D$ band intensity associated with the top edge in Figure \ref{fig1}(d)  (forming a relative angle of $\sim$\,60$^{0}$ with $\vec{P}_{0}$)  is weaker than that obtained from the other edges. Figure \ref{fig1}(e) shows Raman scattering spectra acquired at two different locations [indicated in Fig.~\ref{fig1}(a)]. The upper spectrum was acquired near the edge of the graphene layer whereas the lower spectrum was recorded $\approx 1\,\mu$m from the edge.  The $D$ band appears only in the spectrum acquired near the edge indicating that the graphene sheet is free of structural defects. The Raman scattering spectra also reveal that the $G^{\prime}$ band is composed of a single peak, which confirms that the sample is a {\em single} graphene sheet~\cite{ferrari02,eklund,graf}.

\begin{figure}
\begin{center}
\includegraphics[width=0.38\textwidth]{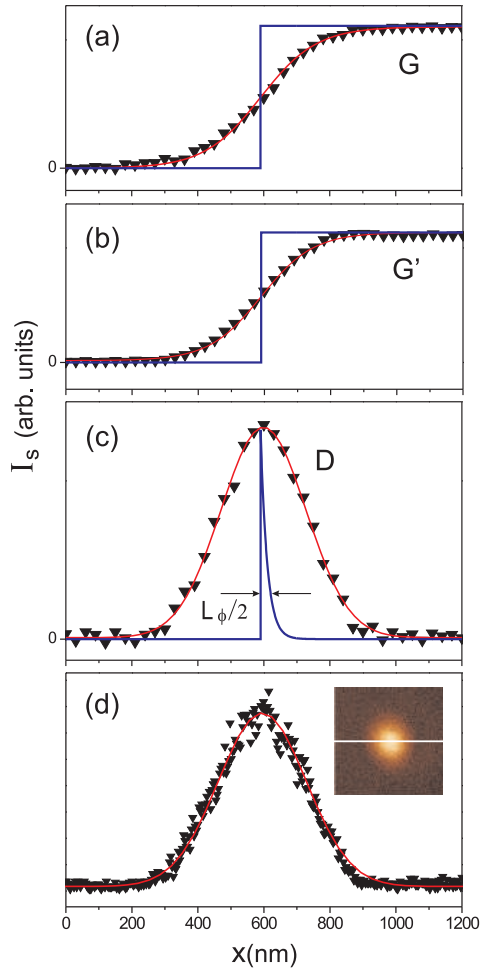}
\caption{(a)-(c): Intensities of the $G$, $G^{\prime}$, and $D$ bands as a function of the relative distance between laser focus and graphene edge. The data has been recorded along the dotted line in Fig.~\ref{fig1}(a). Triangles indicate experimental data and the red curves are reconstructions according to Eq.~(\ref{eq03}) using the Raman susceptibilities $\chi_s$ shown as blue curves.  (d):
Fluorescence intensity profile of a single molecule (nile blue) deposited on the same sample. The inset shows the corresponding confocal fluorescence rate together with the line along which the intensity profile was evaluated.  \label{fig2}}
\end{center}
\end{figure}

In order to analyze the spatial dependence of the $D$, $G$ and $G^{\prime}$ band intensities near the graphene edge, Raman spectra similar to those shown in Figure \ref{fig1}(e) were recorded as the position of the incident laser focus was moved in steps of 30\,nm along a 1.2\,$\mu$m line perpendicular to the edge [dotted line in Figure \ref{fig1}(a)]. Figure \ref{fig2} shows the measurement (dark triangles) of the $G$, $G^{\prime}$, and $D$ band intensities along the selected line. As the graphene edge is moved through the laser focus, the intensities of the $G$ and $G^{\prime}$ bands gradually transit from a minimum value (dark counts) to a maximum value. On the other hand, the $D$ band intensity achieves a maximum value when the graphene edge is in the laser focus. In order to extract the material-specific response functions (Raman susceptibilities $\chi_s$) the Raman intensity measurements have to be deconvolved with the field distribution $\vec{E}({\bf r})$ of the excitation laser (point spread function). This inverse scattering problem is nontrivial and can only be accomplished by multiple sets of measurements (tomography) or by use of prior information.

Because the size of the excitation field is much larger than the electron-phonon scattering length in graphene, the measured Raman intensities can be expressed as incoherent spatial sums, i.e. \begin{equation}
I_s(x,\omega_{s})\propto\,\int^{+\infty}_{-\infty}\!\!\left|\,\tensor{\chi}_s(x^{\prime};\omega_{s},\omega)\,\vec{E}(x^{\prime}\!-\!x,\omega)\right|^{\,2} dx^{\prime} \; ,
\label{eq03}
\end{equation}
where $x$ designates the lateral position of the laser focus and  $s \in \{G,G',D\}$. The $y$-dependence has been eliminated by integration (vertical edge). Using the fact that $\vec{E}$ has negligibly weak polarization components perpendicular to the graphene edge and that the phase distribution in the focal plane is  uniform~\cite{novobook} allows Eq.~(\ref{eq03}) to be expressed in scalar form. The Raman susceptibilities $\chi_s$ represent the local interaction strength between incident light and a specific phonon mode.

In order to solve for $\chi_s$ we require an accurate measurement of the excitation field $\vec{E}$. This task has been accomplished by spin-casting a 1nM solution of nile blue molecules onto the graphene sample and acquiring fluorescence rate images of single molecules with an attenuated laser power ($\approx 200\,$nW). Each molecule maps out the spatial distribution of the excitation intensity ($|\vec{E}|$)~\cite{novobook} and hence renders the excitation profile for Eq.~(\ref{eq03}). Thus, the excitation field $\vec{E}$ is measured under the same experimental conditions as the intensities $I_s$ of the different Raman lines.

Figure \ref{fig2}(d) shows the fluorescence emission intensity of a single molecule along the $x$-direction. The curve corresponds to a line cut through the center of the fluorescence rate image shown in the inset. The red curve is a Gaussian fit according to $\exp[-x^2/\gamma^{2}]$, with $\gamma=186.5\,$nm. It turns out that the fluorescence curve in Fig.~\ref{fig2}(d) is narrower than the profile of the $D$ band intensity shown in Fig.~\ref{fig2}(c), which indicates that the response function $\chi_D$ has finite extent.

In order to proceed we consider the fact that $\chi_s(x)$ vanishes for $x$ smaller than a certain value $x_e$ (the edge) and this value has to be the same for {\em all} Raman lines. According to this criterion we use a one-parameter model for the Raman susceptibilities. For the $G$ and $G^{\prime}$ bands we use
\begin{equation}\label{eq05}
\chi_{G,G^{\prime}} (x)= \left\{
\begin{array}{cc}
 1- \exp[-(x-x_e)/x_{G,G^{\prime}}] & x \geq x_{e}\\
0\; & x<x_{e}
\end{array}\right.\;,
\end{equation}
which we insert into Eq.~(\ref{eq03}), together with the experimentally determined excitation field $\vec{E}$. Numerical integration allows us to find a Raman intensity profile $I_{G,G^{\prime}}$ which can then be compared with the experimental data (dark triangles). Using this procedure, the parameters $x_{G,G^{\prime}}$ and $x_e$ can be solved in a least-squares sense. We find that $x_{G,G^{\prime}}\approx 0$ and hence the susceptibilities $\chi_{G}$ and $\chi_{G'}$ become step functions as indicated by the blue curves in Figs.~\ref{fig2}(a) and (b). The red curves are the calculated fits for $I_{G}$ and $I_{G^{\prime}}$ according to relations (\ref{eq03}) and (\ref{eq05}). The best fit for the edge position turns out to be $x_{e}$\,=\,590\,nm, which is 10\,nm displaced from the maximum of the $D$ band intensity curve (Fig.~\ref{fig2}c) and provides further evidence for the finite width of the $D$ band response function.

For the $D$ band susceptibility we use the following one-parameter model
\begin{equation}
\chi_{D}(x)= \left\{
\begin{array}{cc}
 \exp[-(x-x_e)/x_D]\ & x \geq x_{e}\\
0  & x<x_{e}
\end{array}\right.\;,
\label{eq06}
\end{equation}
and follow the same procedure as for the $G$ and $G'$ bands.  The best fit is found for $x_{D}$\,=\,20\,nm and the calculated susceptibility according to Eq.~(\ref{eq06}) is shown as a blue curve in Fig.~\ref{fig2}(c).  The red curve is a calculated fit for $I_{D}$ according to relations (\ref{eq03}) and (\ref{eq06}). The finite value of $x_{D}$ is the central result of this study and proves that the $D$ band response function is not completely localized to the graphene edge.

We now discuss the origin of the spatial extent of the $D$ band susceptibility at the graphene edge. The double-resonance process giving rise to the $D$ band involves the inelastic scattering of a $\pi$ electron in the conduction band by a TO phonon \cite{ferrpho}, whose wavevector $\vec{q}$\ lies near the vertices of the first Brillouin zone of graphene ($K$ and $K^{\prime}$ points) \cite{thomsen,saito}. Momentum conservation is only satisfied in the scattering process if the electron is elastically back-scattered by a defect providing a wavevector $\vec{d}$\,$\sim$\,-\,$\vec{q}$. In real space, the defect (edge) of our graphene sample is localized in the $x$ direction. Consequently, it is completely delocalized in the reciprocal space along the same direction, thereby providing the necessary condition for momentum conservation in the double-resonance mechanism giving rise to the $D$ band \cite{edge}. The closer an electron is located to the graphene edge the higher is its probability to be involved in $D$ band scattering.

The {\em phase-breaking length} $L_{\phi}$ of a conduction electron is defined as the average distance traveled before undergoing inelastic scattering~\cite{solid01}. Electrons involved in $D$ band Raman scattering undergo a single inelastic scattering event with a lattice phonon. Therefore, $L_{\phi}$ corresponds to the average distance traveled by such electrons during the time interval in which the D band scattering process takes place. Considering that the electron can travel two distances (to the edge and away from it) $L_{\phi}$ is estimated as roughly twice the $1/e$ value of the response function $\chi_D$, that is, $L_{\phi}\,\approx\,40\,$nm [see Figure \ref{fig2}(c)]. Interestingly, this value is smaller than the average value of $\sim$\,100\,nm predicted for micron-sized samples at room temperatures~\cite{mean01,berger02}. We attribute this difference to the fact that the region near the edge presents a higher density of structural defects, causing a reduction of the {\em phase-breaking length}~\cite{mean02}.

Finally, it is worth noting that, since the frequency and wavevectors of phonons involved in the double-resonance process giving rise to the $D$ band are defined by the incident laser excitation~\cite{thomsen,saito}, the {\em phase-breaking length} measured here is associated with a particular location in the phonon dispersion curve. This has clear advantages over standard transport measurements because contributions due to different scattering processes (electron-electron, electron-phonon, and electron-impurity) are not intermixed.

In summary, we have presented an optical measurement of the {\em phase-breaking length} of conduction electrons in graphene based on measuring the spatial dependence of the Raman spectrum near the edge of a single graphene layer. We have reconstructed the $D$ band susceptibility $\chi_D$ near the graphene edge by simultaneous measurements of Raman spectra and the point spread function of the incident laser beam. Based on the theory of double-resonance scattering, we associate the mean length of $\chi_D$  with the {\em phase-breaking length} of conduction electrons due to inelastic scattering with TO phonons near the vertices of the 1$^{st}$ Brillouin zone of graphene. Using this approach we determine $L_{\phi}$\,=\,40\,nm. This result is also valid for holes in the valence band, and shows the influence of the edges on the transport properties of charge carriers in nanoscale graphene systems where edge defects are inevitably present. An interesting extension of this work will be the simultaneous measurement of Stokes and anti-Stokes $D$ band profiles under different temperature conditions. These measurements would reveal the dependence of the {\em phase-breaking length} on the phonon population.

This work was supported by DOE (grant DE-FG02-05ER46207) and NSF (grant CHE-0454704).

\end{document}